\documentclass[12pt]{article}
\usepackage{epsf}
\def\be{\begin{equation}}
\def\ee{\end{equation}}
\def\bea{\begin{eqnarray}}
\def\eea{\end{eqnarray}}

\begin{document}
\begin{titlepage}
\begin{center}
{\Large \bf William I. Fine Theoretical Physics Institute \\
University of Minnesota \\}
\end{center}
\vspace{0.2in}
\begin{flushright}
FTPI-MINN-13/02 \\
UMN-TH-3133/13 \\
January 2013 \\
\end{flushright}
\vspace{0.3in}
\begin{center}
{\Large \bf Enhanced mixing of partial waves near threshold for heavy meson pairs and properties of $Z_b(10610)$ and $Z_b(10650)$ resonances
\\}
\vspace{0.2in}
{\bf M.B. Voloshin  \\ }
William I. Fine Theoretical Physics Institute, University of
Minnesota,\\ Minneapolis, MN 55455, USA \\
School of Physics and Astronomy, University of Minnesota, Minneapolis, MN 55455, USA \\ and \\
Institute of Theoretical and Experimental Physics, Moscow, 117218, Russia
\\[0.2in]

\end{center}

\vspace{0.2in}

\begin{abstract}
The mixing of $S-D$ partial waves for the heavy meson pairs in the decays $\Upsilon(5S) \to [B^* \bar B+ {\rm c.c.}] \pi$ and $\Upsilon(5S) \to B^* \bar B^* \pi$ is considered. It is argued that this mixing, enhanced by the heavy meson mass, is calculable as dominated by a rescattering through pion exchange if the production of the heavy mesons is dominated by $S$-wave `molecular' resonances $Z_b(10610)$ and $Z_b(10650)$. The effect of the mixing grows with the c.m. energy $E$ in each channel over the threshold, and may reach $10 \div 20$\% in the rate at the upper end of applicability of the discussed approach, $E \sim 15\,$MeV. It is also argued that the mixing is likely to reach maximum at energy approximately equal to the mass gap between the thresholds.
\end{abstract}
\end{titlepage}

The strong interaction between hadrons containing heavy and light quarks results in a peculiar dynamics near a threshold for pairs of such hadrons. One of the recent most spectacular observations of this phenomenon is the discovery by Belle~\cite{bellez} of the `twin' isotriplet resonances $Z_b(10610)$ and $Z_b(10650)$ with masses within few MeV from the respective $B^* \bar B$ and $B^* \bar B^*$ thresholds. The resonances were initially observed in the decays $\Upsilon(5S) \to \Upsilon(nS) \pi \pi$ with $n=1,2,3$ and $\Upsilon(5S) \to h_b(mP) \pi \pi$ with $m=1$ and 2 as conspicuous peaks in the invariant mass spectra for the system of bottomonium plus pion.   These two peaks are naturally explained~\cite{bgmmv} as `molecular' states with quantum numbers $I^G(J^P)=1^+(1^+)$ made out of $S$-wave pairs of mesons: $Z_b(10650) \sim B^* \bar B^*$, $Z_b(10610) \sim (B^* \bar B - B \bar B^*)$. This picture agrees well with the subsequent measurement\cite{bellebb} of the invariant mass distribution for the $B^* \bar B  + B \bar B^*$ and $B^* \bar B^*$ meson pairs in the decays $\Upsilon(5S) \to [B^* \bar B+ {\rm c.c.}]^\pm \pi^\mp$ and $\Upsilon(5S) \to [B^* \bar B^*]^\pm \pi^\mp$. The distribution in each channel displays a strong peak near the respective threshold, and in fact the data, although still with a large uncertainty, suggest that the entire distribution is dominated by the corresponding $Z_b$ resonance. 

Naturally, more precise data, if available in the future, would allow a better analysis of the internal dynamics of the $Z_b$ resonances. Such analysis however would require a better understanding of finer effects, including those in the spectra of the heavy meson pairs near the threshold resonances. The goal of this paper is to point out that the spectra in the vicinity of the resonances can be noticeably affected by an enhanced mixing of partial waves, specifically the $S-D$ mixing, due to rescattering of the heavy mesons. In particular, in the (likely) case that the $Z_b$ resonances are indeed very strongly dominated by the $S$-wave heavy meson pairs, a measurable $D$-wave should arise due to rescattering at energies starting from the excitation energy above $\sim 10$\, MeV over the threshold\footnote{The effect of the $S-D$ mixing due to rescattering in the formation of the $Z_b$ resonances as `molecular' bound states was recently considered in Refs.~\cite{shllz,lslz} with the conclusion that this effect is small in the wave function of the meson pair in the bound state.}. The onset of the $D$-wave in the spectrum is dominated by the pion exchange between the heavy mesons and is, to an extent, calculable. At higher excitation energy, however, the presence of the $D$-wave can only be estimated parametrically. The presence of the $D$ wave can be measured in future experiments by studying the angular distribution of the heavy meson pairs and it is likely to be important for a better determination of the parameters of the $Z_b$ resonances.

The enhancement of the effects of the interaction of heavy hadrons through the light degrees of freedom is well known. Indeed, in the limit of large mass $M$ of the heavy quark the latter interaction, described by a potential $V$, does not depend on $M$. Thus at a given momentum scale $p$ any effect of the interaction enters through the product $M V$ and thus gets bigger at large $M$. In particular, if $V$ describes the spin-dependent interaction resulting in an $S-D$ mixing, the effect of the mixing should generally contain the enhancement factor $M/\Lambda_{QCD}$ in the amplitude. In a situation where the strength and the range of the potential are determined by $\Lambda_{QCD}$, an $S$-wave state rescatters into a $D$-wave state with the ratio of the amplitudes generally estimated as
\be
{A_D \over A_S} \propto {M \, p^2 \over \Lambda_{QCD}^3}~,
\label{adas}
\ee
where $p$ is the c.m. momentum of the heavy hadrons, and the estimate applies as long as $p^2 \ll \Lambda_{QCD}^2$. This estimate, however, should be further modified for the effect of the pion exchange, since the pion mass $\mu$ can be considered as small in the scale of $\Lambda_{QCD}$. The contribution of the pion exchange to the wave mixing can be estimated as
\be
{A_D \over A_S} \propto g^2 \, {M \, p^2 \over \Lambda_{QCD} \, \mu^2} ~~~~~{\rm at}~ p^2 \ll \mu^2~,
\label{lowp}
\ee
and
\be
{A_D \over A_S} \propto g^2 \, {M  \over \Lambda_{QCD} } ~~~~~{\rm at}~ \Lambda_{QCD}^2 \gg p^2 \gg \mu^2~, 
\label{hip}
\ee
where $g$ is the constant for pion interaction with heavy hadrons. Clearly, these estimates imply that the pion exchange dominates the wave mixing at the energies corresponding to $p^2 \ll \Lambda_{QCD}^2$ and becomes comparable with the effect of other contributions to the spin-dependent interaction at the upper end of the applicability of both estimates (\ref{adas}) and (\ref{hip}) where the pion exchange can no longer be separated from those other short distance contributions. Neither the effect of the pion exchange at such momenta is calculable due to unknown form factors in both the pion interaction with heavy hadrons and in the short-distance behavior of $A_S$. For this reason the specific calculations in this paper are limited to the energies above the threshold corresponding to $p^2 \ll \Lambda_{QCD}^2$. Although restricted to this low energy range the calculation to be presented indicates that the arising from the rescattering $D$ wave can be well measurable just above the $Z_b$ peaks, corresponding to more than 10\% in the rate at $p \approx 300$\,MeV, i.e. at the c.m. energy $E \approx 17$\,MeV. Due to a moderate magnitude of the mixing one can neglect the back reaction, i.e. the feedback from the $D$ wave to the $S$-wave, which approximation is assumed in the rest of this paper.

For the specific calculation in the discussed processes $\Upsilon(5S) \to [B^* \bar B+ {\rm c.c.}] \pi$ and $\Upsilon(5S) \to B^* \bar B^* \pi$ one can write the pion interaction with the $B$ and $B^*$ mesons in the form (see e.g. in Ref.\cite{mp})
\be
H_{int} = {g \over f_\pi} \, \left \{ \left [ (V^\dagger_l \tau^a P) + {\rm h.c.} \right ]+ i \, \epsilon_{ljk} \, (V^\dagger_j \tau^a V_k) \right \}  \,  \partial_l \pi^a
\label{lint}
\ee
with $V_i$ and $P$ standing for the wave functions (in the nonrelativistic normalization) of the vector $(B^*)$ and the pseudoscalar $(B)$ meson isotopic doublets and $\tau^a$ being the isospin Pauli matrices. The constant $f_\pi \approx 132\,$MeV is used in Eq.(\ref{lint}) for normalization. The dimensionless pion coupling $g$ can then be evaluated by using the heavy quark symmetry and the known\cite{pdg} rate of the $D^{*+} \to D \pi$ decay: $g^2 \approx 0.18$.

The mixing of the partial waves due to the pion arising from the `diagonal' rescattering in the channels $B^* \bar B$ and $B^* \bar B^*$ is shown in the Figure 1. Only these `diagonal' processes receive an unsuppressed contribution from the domain of the loop momentum $\vec q$ such that $q^2 \ll \Lambda_{QCD}^2$ provided that the overall c.m. momentum $p$ of the mesons satisfies the same condition. In particular, this is not the case for the rescattering between the two channels: $B^* \bar B \leftrightarrow B^* \bar B^*$, where due to the mass difference $\Delta M$ between the $B^*$ and $B$ mesons the minimal momentum transfer through the pion propagator is of order $q^2 \sim M \, \Delta M$, which corresponds to $q^2 \sim \Lambda_{QCD}^2$ both numerically and parametrically. Thus the non-diagonal rescattering produces a mixing effect only of the order described by Eq.(\ref{adas}). 
\begin{figure}[ht]
\begin{center}
 \leavevmode
    \epsfxsize=16cm
    \epsfbox{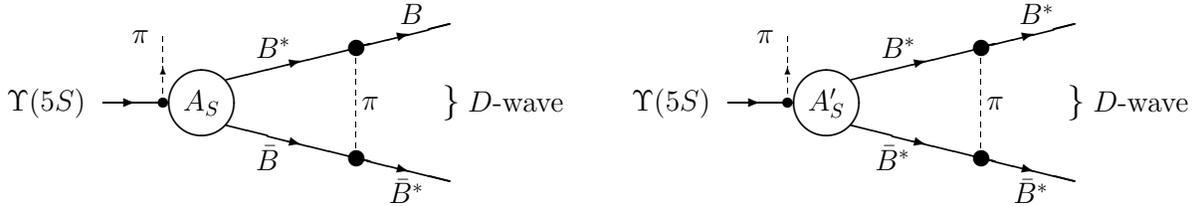}
    \caption{The rescattering through the pion exchange resulting in an $S - D$ mixing in the channels $B^* \bar B$ and $B^* \bar B^*$.)}
\end{center}
\end{figure} 

Using the notation $\vec a$ and $\vec b$ for the polarization amplitudes of the outgoing $B^*$ and $\bar B^*$ mesons, one can write the general form of the $\Upsilon(5S)$ decay amplitudes:
\bea
&&A[\Upsilon(5S) \to [B^* \bar B+ {\rm c.c.}] \pi]/E_\pi=  {\Upsilon_i (a_j-b_j) \over \sqrt{2}} \, \left [ A_S \, \delta_{ij} + {1 \over \sqrt{2}} \, A_D \, (3 \, n_i n_j - \delta_{ij}) \right ]~, \nonumber \\
&&A[\Upsilon(5S) \to B^* \bar B^* \pi]/E_\pi=  {\Upsilon_i \epsilon_{j k l} a_k b_l\over \sqrt{2}} \, \left [ A'_S \, \delta_{ij} + {1 \over \sqrt{2}} \, A'_D \, (3 \, n_i n_j - \delta_{ij}) \right ]~,
\label{amps}
\eea
where $\vec n = \vec p/p$ is the unit vector in the direction of c.m. momentum of one of the mesons, and $p=|\vec p|$. The factor $1/E_\pi$ in l.h.s. removes from the amplitude the chiral dependence on the energy $E_\pi$ of the emitted pion, which dependence is not the subject of the discussion in this paper. The amplitudes $A_S$ and $A_D$ are functions of $p^2$ and their relative normalization  is chosen in such a way that the probability is proportional to the sum of their squares: $|A_S|^2 + |A_D|^2$. Under these notational conventions the amplitude of the $D$ wave generated by the graphs of Fig.~1 can be calculated by means of nonrelativistic perturbation theory as
\bea
&&A^{(')}_D (p^2)  = {g^2 \over \sqrt{2} \, f_\pi^2} \, \int \, A^{(')}_S(q^2) \,  {M \over q^2-p^2- i \epsilon} \, \left ( n_i n_j - {1 \over 3} \, \delta_{ij} \right ) \, {(q_i-p_i) (q_j-p_j) \over (\vec q - \vec p)^2 + \mu^2 } \, {d^3 q \over (2 \pi)^3} = \nonumber \\
&& {g^2  \over 48 \, \pi^2 \, \sqrt{2} \, f_\pi^2} \, \int_0^\infty  \, A^{(')}_S(q^2) \, {M \, q \over q^2-p^2- i \epsilon} F(p,q) dq^2
\label{intamp}
\eea
where in the latter transition an averaging over the relative angle between the vectors $\vec q$ and $\vec p$ is performed, so that only an integration over the scalar variable $q^2$ remains with the weight function $F(p,q)$ given by
\be
F(p,q) = {5 p^2 - 3 q^2 - 3 \mu^2 \over p^2}+ {4 \mu^2 \, p^2 + 3 (p^2-q^2-\mu^2)^2 \over 4 \, p^3 \, q} \, \ln {(p+q)^2 + \mu^2 \over (p-q)^2 + \mu^2}~.
\label{fpq}
\ee
In Eq.(\ref{intamp}) the notation $A^{(')}$ signifies that the relation can be used in either of the two heavy meson channels, and $M$ stands for two times the reduced mass of the two mesons. For the purpose of the present calculation the difference in this reduced mass in the two channels can be neglected and a common value $M \approx 5.3$\,GeV used.

The absorptive part of the ratio of the $D$-wave and $S$-wave amplitudes, corresponding to the on-shell cut of the graphs of Fig.~1 can now be found unambiguously by setting $q=p$ in the integrand in Eq.(\ref{intamp}):
\be
r \equiv {\rm Im} \left . {A^{(')}_D \over A^{(')}_S} \right |_{\rm abs} = {g^2  \over 48 \, \pi \, \sqrt{2} } {M \, p \over f_\pi^2} \, \left [ 2- 3 \, {\mu^2 \over p^2} + \left ( {\mu^2 \over p^2} + {3 \over 4} \, {\mu^4 \over p^4} \right ) \, \ln \left ( 1+ 4 \, {p^2 \over \mu^2} \right ) \right ]~.
\label{absor}
\ee
Clearly, this expression exhibits the behavior described by Eqs.(\ref{lowp}) and (\ref{hip}) with one extra power of $p/\Lambda_{QCD}$ arising from the phase space in the intermediate state. The plot of the ratio $r$ given by Eq.(\ref{absor}) at $g^2=0.18$ is shown in Fig.~2 and illustrates the magnitude of the discussed effect of the $S-D$ mixing.
\begin{figure}[ht]
\begin{center}
 \leavevmode
    \epsfxsize=10cm
    \epsfbox{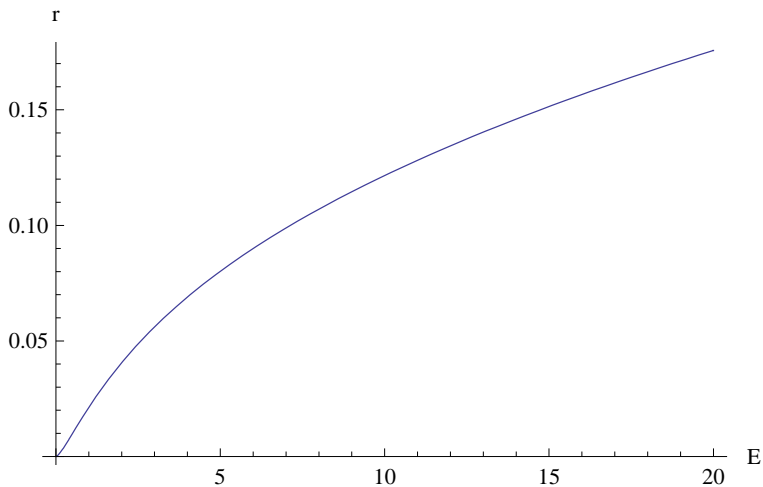}
    \caption{The ratio $r$ described by Eq.(\ref{absor}) as a function of the excitation energy above the threshold $E=p^2/M$.}
\end{center}
\end{figure} 

One should notice that the expression (\ref{absor}) does not describe all of the imaginary part of the ratio of the amplitudes and thus cannot be used to establish a lower bound on the $D$-wave generated through rescattering. The reason is that the amplitude $A^{(')}_S(q^2)$ itself is complex with a phase generally depending on $q^2$. In other words the graphs of Fig.~1 also generally have nontrivial on-shell cuts across the bubble depicting the $S$-wave amplitude. In order to evaluate the full generated $D$-wave amplitude a knowledge of the energy behavior of the $S$-wave amplitude is needed. The data~\cite{bellebb} indicate that the $Z_b$ resonances dominate the spectrum of the heavy meson pairs in the discussed decays of $\Upsilon(5S)$. Thus it appears  reasonable to use in Eq.(\ref{intamp}) a pure resonance expression 
\be
A_S(p^2) = {c \over E - E_0 + i \Gamma/2}= { {\rm const} \over p^2 - p_0^2 + i M \Gamma/2}~,
\label{resa}
\ee
with $E_0$ being the energy position of the resonance relative to the threshold and $p_0^2 = M E_0$. According to the current data~\cite{pdg},  the value of $E_0$ is $2.7 \pm 2.0\,$MeV for $Z_b(10610)$ and $1.8 \pm 1.5\,$MeV for $Z_b(10650)$, while the corresponding values of $\Gamma$ are $18.4 \pm 2.4\,$MeV and $11.5 \pm 2.2\,$MeV. Clearly, using the expression (\ref{resa}) for $A_S$ makes the integral in Eq.(\ref{intamp}) convergent and determined by a scale, which is a combination of $p^2$, $p_0^2$, $M \Gamma/2$ and $\mu^2$, all of which are assumed to be small in the scale of $\Lambda_{QCD}$ and which justifies using the pion exchange amplitude as in Eq.(\ref{intamp}) without introducing a form factor. The integral can then be readily calculated, and the plots in Fig.3 illustrate the magnitude of the discussed generated $D$-wave. For this illustration the value $E_0=0$ is used and the effect depends very weakly on this particular value as long as $E_0$ is smaller than $\Gamma/2$. Also for this illustration the value $\Gamma = 14\,$MeV is used, which is close to the data for both $Z_b$ resonances. One can see that at the upper end of applicability of the discussed approach, i.e. at $E \approx 15 \div 17$\,MeV the relative contribution of the generated $D$ wave can amount to $10 \div 20$\% in the event rate.
\begin{figure}[ht]
\begin{center}
 \leavevmode
    \epsfxsize=10cm
    \epsfbox{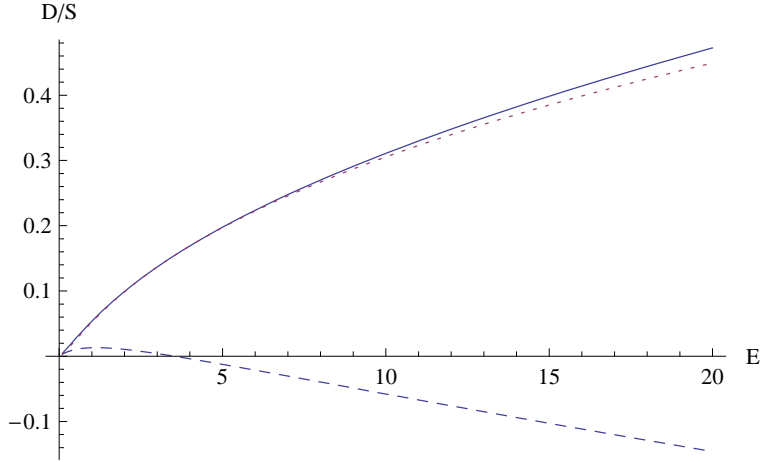}
    \caption{The absolute value (solid) and the real (dashed) and imaginary (dotted) parts of the ratio $A_D/A_S$ for the $D$-wave amplitude generated from the $A_S$ given by Eq.(\ref{resa}) with $E_0=0$ and $\Gamma=14\,$MeV.}
\end{center}
\end{figure} 

It can be noticed that the simplest Breit-Wigner formula for the resonance amplitude is used for $A_S$ in Eq.(\ref{resa}). In particular the width parameter $\Gamma$ is assumed to be constant and does not include the energy-dependent absorption in the heavy meson channel above the threshold. This approximation appears to be sufficient with current data, but may need to be modified in a thorough analysis of possible more detailed data in the future.

As one can see from the plots of Fig.~3, the $S-D$ mixing is quite small at low energies $E \sim 1\,$MeV, which are relevant for the internal dynamics of the $Z_b$ resonances, so that the mixing is likely only a minor effect in this internal dynamics, in agreement with the conclusions of Refs.~\cite{shllz,lslz}. The ratio $A_D/A_S$ however grows with the excitation energy and can possibly be studied in the decays from $\Upsilon(5S)$.
At present there seem to be no theoretical means of analyzing the behavior of the $S-D$ mixing at larger excitation energies $E$ beyond the region of applicability of the presented approach, i.e. at $E$ being not small as compared to $\Lambda_{QCD}^2/M$. It is however interesting to notice that if higher values of $E$  were accessible in the $\Upsilon(5S)$ decay, i.e. in an artificial theoretical limit of a large mass of $\Upsilon(5S)$, the $S-D$ mixing should vanish at high $E$. Indeed, the strong interaction rescattering arises from the interaction of the light quarks in the heavy mesons, and the spin of the heavy $b$ quarks is conserved, and the whole process can be considered with spinless $b$ quarks. With spinless $b$ quarks the $\Upsilon(5S)$ would be an $J^P=0^+$ state, and after emission of an $S$-wave pion would produce heavy meson pair with quantum numbers $J^P=0^-$, for which obviously no $S-D$  mixing is possible. In terms of the mesons, the channels $B^* \bar B$ and $B^* \bar B^*$ would degenerate into one channel as well as the two $Z_b$ resonances would coalesce into one. The vanishing of the $S-D$ mixing then occurs through a cancellation between the `diagonal' rescattering, e.g. $B^* \bar B^* \to B^* \bar B^*$ and the `off-diagonal' $B^* \bar B \to B^* \bar B^*$. For the actual heavy mesons, whose thresholds are split by $\Delta M$, in the region of dominance of the pion exchange only the `diagonal' scattering should be retained, as discussed previously. However at large $E$, specifically at $E \gg \Delta M$ the difference in the thresholds becomes unimportant and the heavy quark spin symmetry behavior should set in. Therefore one can reasonably expect that the $D/S$ ratio has a maximum at $E \sim \Delta M$ and decreases at higher energy. It would be quite interesting if this behavior can be studied in the decays from the actual $\Upsilon(5S)$.

It can be also mentioned that a similar enhanced mixing of partial waves, namely the $P-F$ mixing, can be expected in the production of $B^* \bar B^*$ pairs in $e^+e^-$ annihilation at the c.m. energy just above the threshold. A detailed discussion of the effects of this mixing will be reported separately.

This work is supported, in part, by the DOE grant DE-FG02-94ER40823.

\end{document}